\pdfoutput=1
\documentclass[journal,article,submit,pdftex,moreauthors]{mdpi}

\firstpage{1} 
\pubvolume{1}
\issuenum{1}
\articlenumber{0}
\pubyear{2023}
\copyrightyear{2023}
\datereceived{ } 
\daterevised{ } % Comment out if no revised date
\dateaccepted{ } 
\datepublished{ } 
\hreflink{https://doi.org/} % If needed use \linebreak
\pdfoutput=1 % Uncommented for upload to arXiv.org
\DeclareUnicodeCharacter{2212}{\textendash}

\Title{Polarization sensitivity in scattering-type \\scanning near-field optical microscopy \\ -- -- \\towards nanoellipsometry}

\TitleCitation{Title}

\Author{Felix G. Kaps $^{1,2}$ , Susanne C. Kehr $^{1}$*, and Lukas M. Eng $^{1,2}$}

\AuthorNames{Felix G. Kaps, Susanne C. Kehr and Lukas M. Eng}

\AuthorCitation{Kaps, F.G.; Kehr, S.; Eng, L.}
\address{%
$^{1}$ \quad Institute of Applied Physics, TUD Dresden University of Technology, 01062 Dresden, Germany\\
$^{2}$ \quad Würzburg-Dresden Cluster of Excellence - EXC 2147 (ct.qmat), TUD Dresden University of Technology, 01062 Dresden, Germany}

\corres{Correspondence: susanne.kehr@tu-dresden.de}

\abstract{Electric field enhancement mediated through sharp tips in scattering-type scanning near-field optical microscopy (s-SNOM) enables optical material analysis down to the 10-nm length scale, and even below. Nevertheless, mostly the out-of-plane electric field component is considered here due to the lightning rod effect of the elongated s-SNOM tip being orders of magnitude stronger as compared to any in-plane field component. Nonetheless,  the fundamental understanding of resonantly excited near-field coupled systems clearly allows us to take profit from all vectorial components, especially also from the in-plane ones. In this paper, we theoretically and experimentally explore how linear polarization control of both near-field illumination and detection, can constructively be implemented to (non-)resonantly couple to selected sample permittivity tensor components, e.g. explicitly also to the in-plane directions. When applying the point-dipole model, we show that resonantly excited samples respond with a strong near-field signal, to all linear polarization angles. We then experimentally investigate the polarization-dependent responses for both non-resonant (Au) and phonon-resonant (3C-SiC) sample excitations at a 10.6~µm and 10.7~µm incident wavelength using a tabletop CO$_2$ laser. Varying the illumination polarization angle thus allows for quantitatively comparing the scattered near-field signatures for the two wavelengths. Finally, we compare our experimental data to simulation results, and thus gain the fundamental understanding of the polarization’s influence on the near-field interaction. As a result, the near-field components parallel and perpendicular to the sample surface can be easily disentangled and quantified through their polarization signatures, connecting them directly to the sample's local permittivity.}

\keyword{{Nanooptics; materials science; nanotechnology; nanoellipsometry; s-polarization; \\p-polarization; s-SNOM; resonant; non-resonant.}}

\begin{document}

\section{Introduction}

Since being a versatile tool for optical surface characterization and analysis, scattering-type scanning near-field optical microscopy (s-SNOM) has been broadly applied to various material systems, ranging from single crystals \cite{Rao2020}, thin films \cite{Tesema2022}, heterostructures \cite{Dai2019}, semiconductors \cite{Ritchie2022}, to biological samples \cite{Rygula2018}. Generally, s-SNOM can be performed at any wavelength \cite{Zenhausern1995, Knoll1999a, Hillenbrand2001, Taubner2003, Huber2008, Kuschewski2016}, and thus the near-field response strongly depends on the sample's optical characteristics, i.e. the frequency-dependent polarizability. Moreover, s-SNOM may couple to the different material properties, such as plasmonic/phononic responses \cite{DeOliveira2021}, crystalline structure \cite{Huber2006}, spins \cite{Cui2021}, and charge carrier concentrations \cite{Huber2006, Huber2008}, with the possibility to quantify all of them via the local optical sample response. Connecting the near-field signal to the local permittivity, therefore, is the uttermost goal in s-SNOM \cite{Kehr2008, Govyadinov2014}. 

Sensitivity to the local in-plane anisotropies has been proven using aperture-type scanning near-field optical microscopy (a-SNOM) \cite{Lee2006}. Moreover, a-SNOM-based nano-ellipsometry was successfully implemented into transmission \cite{Liu2010} and reflection set-ups \cite{Liu2013a}, thus directly connecting the optical responses to the local permittivity tensor elements. So far thin films have been in the focus of those studies that then mostly were inspected in transmission \cite{Liu2010, Liu2013a}.

In comparison, s-SNOM should support a similar in-plane sensitivity as a-SNOM, and moreover offers a much higher and wavelength-independent lateral resolution, since basing on light scattering off a standard scanning force microscopy (SFM) tip. s-SNOM was shown to profit from polarization-sensitive measurements, like mapping plasmonic structures using crossed polarizers \cite{Esslinger2012}, polarization-dependent absorption in organic materials \cite{Mrkyvkova2021}, and probing in-plane anisotropies \cite{Schneider2007}. The challenge now lies in extracting the in-plane response as it is commonly assumed to be suppressed due to the enhanced out-of-plane antenna response of the metal-like SFM tip \cite{McArdle2020}.

Various efforts have been made to enhance the in-plane response of standard s-SNOM probes through, for example, optimizing the AFM tip shape \cite{Ren2021}, or tilt the tip cone with respect to the sample surface normal \cite{Park2018a}. Similarly, the sample's in-plane response can be enhanced by patterning dedicated nano-antennas onto the sample surface \cite{Yao2021}. Moreover, it has been demonstrated both theoretically \cite{Aminpour2020} and experimentally \cite{Wehmeier2020}, that excitation close to the sample's optical resonance, results in an in-plane signal of significant strength even when using standard s-SNOM tips. As this in-plane response is strongly connected to the resonant excitation, its sensitivity to the local sample properties is enhanced for that frequency interval. Hence, minute changes of the local sample properties related to e.g. mechanical stress \cite{Huber2006} and permittivity anisotropies \cite{Schneider2007, Wehmeier2019, Wehmeier2020, Doring2018} are expected to have a drastic impact on the in-plane contributions in s-SNOM.  

This paper explores the possibility of vectorial field analysis under resonant near-field excitation. Notably, this is achieved by rotating and analyzing the linear polarization of both incident and detected fields, respectively, thus disentangling the mixed signal contributions of in- and out-of-plane components. Polarization-sensitive measurements are performed on both 3C-SiC and Au at wavelengths of 10.6~µm and 10.7~µm. Measurements are compared to simulations that combine the point-dipole model \cite{Knoll1999a} and the Jones matrix formalism \cite{Jones1941a} to account for any polarization modification in the detection path. We obtain an excellent fit between theory and experiment, that finally allows us to calculate local permittivity values from the sample's polarization point-spectra. \\

%%%%%%%%%%%%%%%%%%%%%%%%%%%%%%%%%%%%%%%%%%
\section{Theory}
\label{sec:Theory}

% motivation + route: warum dieses model vergleiche mit anderern
Investigating the near-field interaction as a function of polarization requires a theoretical model that supports the calculation of all vector components, i.e. both in- and out-of-plane. The point-dipole model \cite{Knoll1999a} is uniquely able to accomplish this task and, in a first step, is applied here to evaluate the sample's near-field response for the different linear polarization angles of illumination. Next, the resulting field scattered off the tip-sample junction into the far field is simulated using the radiating dipole model \cite{Lukosz1979}, and the light intensity at the detector position is readily computed when accounting for all electric field components in the detection path. Notably this signal contains both a far- and near-field dependence, as will be shown later. Disentangling these components and correlating them to the sample's local permittivity thus is exactly the goal of this paper.

\begin{figure*}[t]
    \centering
    \includegraphics[width = 1\textwidth]{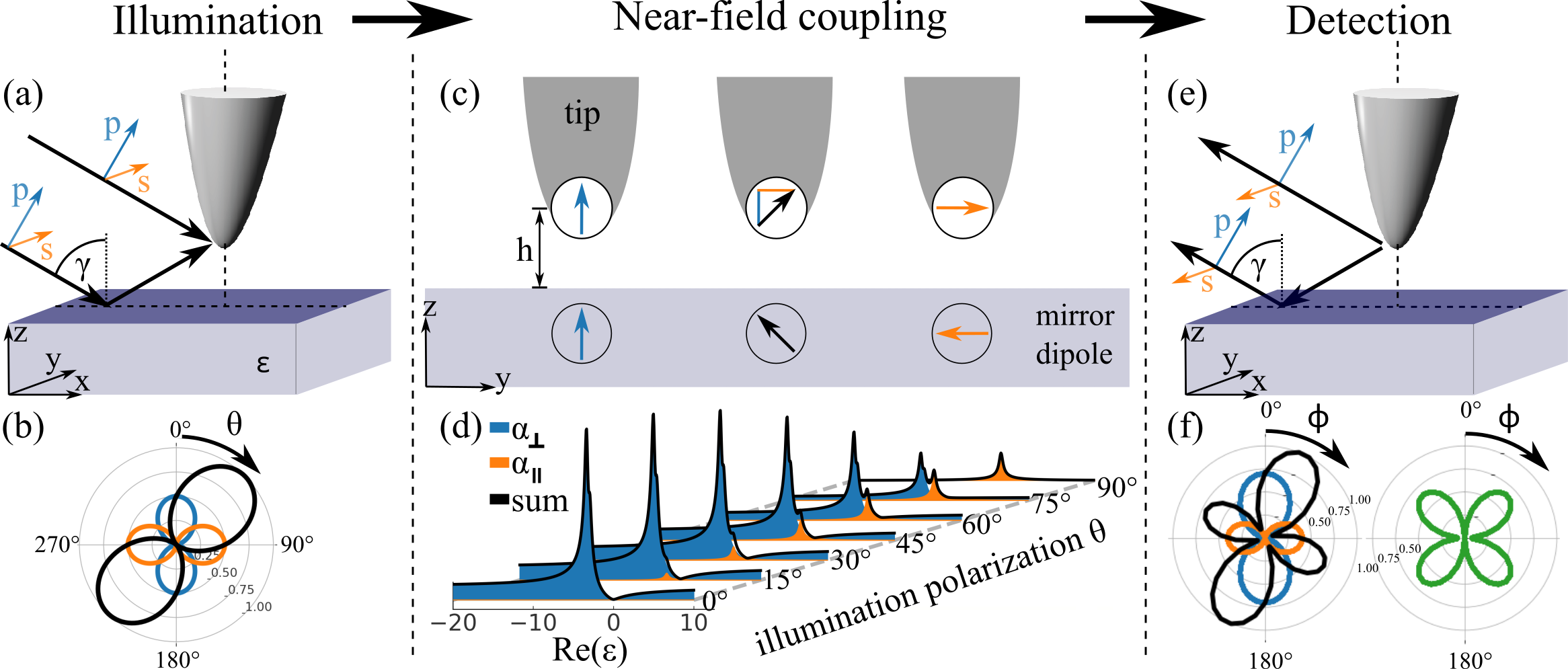}
    \caption{(a) Light polarization geometry of a standard s-SNOM tip under illumination. (b) Linear illumination polarization state with polarization angle $\theta$ = 45$^{\circ}$, separated into individual 'p' (\textcolor{blue}{blue}) and 's' (\textcolor{orange}{orange}) polarized contributions, according to the Jones matrix formalism \cite{Jones1941a}. (c) Interaction between the tip dipole and image dipole according to the point-dipole model for 'p', $45^{\circ}$, and 's' polarized excitation, respectively. (d) Polarizability tensor elements as a function of $\theta$ and Re($\varepsilon$) (with $Im(\varepsilon$) = 0.2, $\gamma$ = 60$^\circ$) highlighting the in-plane and out-of-plane resonances of the near-field coupling at $Re(\varepsilon_{res, \parallel})=-1.72$ and $Re(\varepsilon_{res, \perp})=-3.35$, respectively. (e) Backscattering geometry set-up. (f) The complex polarization state of the backscattered signal showing the 'p' (\textcolor{blue}{blue}), 's' (\textcolor{orange}{orange}), and mixed (\textcolor{green}{green}) contributions. Quantifying this distribution is the goal of this paper here.}
    \label{fig:Setup_and_Permittivity}
\end{figure*}

Figure \ref{fig:Setup_and_Permittivity} (a) sketches the tip-sample system and illumination geometry as used for the study here, both theoretically and experimentally. The incident electric field is described as:

\begin{equation}
    \vec{E}_{in}
    = 
    \begin{pmatrix}
        E_x \\
        E_y \\
        E_z
    \end{pmatrix}
    =
    \begin{pmatrix}
        cos(\gamma) cos(\theta) \\
        sin(\theta) \\
        sin(\gamma) cos(\theta)
    \end{pmatrix}
    \cdot
    \begin{pmatrix}
        1 - r_p \\
        1 + r_s \\
        1 + r_p
    \end{pmatrix}
    \cdot 
     \vec{E_0}
  \hspace{2 mm},
\label{eq:local_elec_field}
\end{equation}

\noindent with $\theta$ the linear polarization angle of the incident beam ($\theta = 0^{\circ}$ corresponds to p-polarization), $\gamma$ the incident angle relative to the sample normal, and $r_s$ and $r_p$ the polarization-dependent Fresnel reflection coefficients. Note that any excitation polarization (except for strict s-polarization with $\theta = 90^{\circ}$) results in both an in- and out-of-plane electric field excitation at the tip-sample junction. Fig \ref{fig:Setup_and_Permittivity} (b) illustrates such a state for $\theta = 45^{\circ}$ indicating both p- (\textcolor{blue}{blue}) and s-polarized (\textcolor{orange}{orange}) components (see also discussion below).

To calculate the near-field interaction of $\vec{E}_{in}$ with an ideal isotropic sample of permittivity $\varepsilon$, the point-dipole model is applied \cite{Knoll2000}. We define the tip polarizability assuming a spherical tip of radius $a$, as $\alpha_{tip} = 4\pi a^3$. It follows:

\begin{equation}
    \vec{P}_{tot } \sim 
     \hat{\alpha}_{tot} \vec{E}_{in}
     =    
    \begin{pmatrix} 
        \alpha_x(\varepsilon, h) & 0 & 0 \\
        0 & \alpha_y(\varepsilon, h) & 0 \\
        0 & 0 & \alpha_z(\varepsilon, h)
    \end{pmatrix} \cdot \begin{pmatrix}
           E_{x} \\
           E_{y} \\
           E_{z} 
    \end{pmatrix}
    \hspace{2 mm},
\label{eq:scattering_cross_section_3D}
\end{equation}

\noindent with $\alpha_i$ the tensor elements of the tip-sample-coupled polarizability tensor, $\varepsilon$ the sample permittivity, and \textit{h} the tip-dipole to sample distance. Using eq. \eqref{eq:scattering_cross_section_3D}, we can separate the near-field coupling into perpendicular (left, \textcolor{blue}{blue}) and parallel (right, \textcolor{orange}{orange}) oriented dipole components as depicted in Figure \ref{fig:Setup_and_Permittivity} (c). The corresponding image dipoles hence will be purely perpendicular and parallel oriented as well. For non-resonant excitation of that spherical tip positioned above an isotropic sample, the different contributions to $\hat{\alpha}_{tot}$ are given as \cite{Knoll2000}:

\begin{equation}
    \alpha_{z}(\varepsilon, h) = \alpha_{\perp}(\varepsilon, h) = \frac{(1+\beta)\alpha_{tip}}{1 - \frac{\beta\alpha_{tip}}{16\pi h^3}}
    \hspace{2 mm},
\label{eq:alpha_z}
\end{equation}

\begin{equation}
    \alpha_{x}(\varepsilon, h) = \alpha_{y}(\varepsilon, h) = \alpha_{\parallel}(\varepsilon, h) = \frac{(1-\beta)\alpha_{tip}}{1 - \frac{\beta\alpha_{tip}}{32\pi h^3}}
    \hspace{2 mm},
\label{eq:alpha_y}
\end{equation}

\noindent with $\beta = \frac{\varepsilon-1}{\varepsilon+1}$ the sample response function, and $\alpha_{\perp}$ and $\alpha_{\parallel}$ referring to the out-of-plane and in-plane total polarizability components with respect to the sample surface, respectively. The calculation becomes more complex but still doable, when considering polarization angles others than $\theta = 0^{\circ}, 90^{\circ}$, since the electric field then is interacting with both $\alpha_{\perp}$ and $\alpha_{\parallel}$ at the same time. Such a scenario is illustrated in \ Fig. \ref{fig:Setup_and_Permittivity} (c), center.

Fig. \ref{fig:Setup_and_Permittivity} (d) displays the mixing of $\alpha_{\perp}$ and $\alpha_{\parallel}$, as calculated with the s-SNOM tip being in firm contact to a fictive sample surface, characterized by $Im(\varepsilon$) = 0.2 = const. for all wavelengths, and the tip-dipole-to-sample distance being $h$ = 0.785$a$. The latter accounts for a shift of the effective tip-dipole position towards the sample upon near-field interaction, as well as for an effective tip-radius of $a$ = 600~nm according to literature \cite{Renger2005, Wehmeier2019}. Overall, the out-of-plane response ($\alpha_{\perp}$) is clearly stronger than the in-plane response ($\alpha_{\parallel}$). Nevertheless, for certain incident polarization angles (e.g. $\theta=75^\circ$), the out-of-plane and in-plane components may reach values that are on the same order of magnitude. Most importantly, both components show resonance peaks at slightly different $Re(\varepsilon)$, here, $Re(\varepsilon_{res, \perp}) = -3.35$ and $Re(\varepsilon_{res, \parallel}) = -1.72$, with their exact positions depending on the tip-sample distance $h$. Since $\varepsilon$ is frequency dependent, spectrally addressing and separating these peaks into in-plane and out-of-plane sample responses becomes possible.

In order to determine the detectablilty of the local in-plane signals in the far field, the radiation of the point dipole $\vec{E}_{near}$ is calculated in the backscattering geometry used here [see Fig. \ref{fig:Setup_and_Permittivity} (e)]. Moreover, additional far-field contributions $\vec{E}_{far}$, e.g. due to direct backscattering, need to be considered, which are superposed to the near-field contribution, resulting in the total scattered signal in Jones vector notation, as:

\begin{equation}
    \vec{E}_{sca} = \vec{E}_{far} + \vec{E}_{near} =
    \begin{pmatrix}
        p_{far} \\
        s_{far}
    \end{pmatrix}
    +
    \begin{pmatrix}
        p_{near} \\
        s_{near}
    \end{pmatrix}
    \hspace{2 mm},
    \label{eq:mixed_signal}
\end{equation}

\noindent with $p_{far}$, $s_{far}$ and $p_{near}$, $s_{near}$ describing the p- and s-polarized far- and near-field components, respectively. Note that in this model, we discuss self-homodyne detection by connecting $\vec{E}_{far}$ to the reflectivity of the sample. Nevertheless, the method is directly applicable to other detection schemes as well, by adapting $\vec{E}_{far}$ accordingly.

Investigating the polarization signature of the scattered signal at the detector ultimately requires to consider all optical elements in the detection path as well. For this purpose, the Jones matrix formalism \cite{Jones1941a} was applied representing every element (i.e. mirrors, beam splitters, etc.) in the detection path by a matrix. With that, the polarization signature $\vec{E}_{det}$ after the detection apparatus for the self-homodyne case is described as:

\begin{equation}
    \vec{E}_{det} = \hat{M}_{mylar} \cdot \hat{M}_{mirror} \cdot \vec{E}_{sca}
    = \vec{E}^{'}_{far} + \vec{E}^{'}_{near}
    = \begin{pmatrix}
        p^{'}_{far} \\
        s^{'}_{far}
    \end{pmatrix}
    +
    \begin{pmatrix}
        p^{'}_{near} \\
        s^{'}_{near}
    \end{pmatrix}
    \hspace{2 mm},
    \label{eq:Jones_detection_path}
\end{equation}

\noindent with $\hat{M}_{mylar}$ \cite{Fymat1971} and $\hat{M}_{mirror}$ being the Jones matrices for the mylar beam splitter and mirror, respectively [see experimental setup in Fig. \ref{fig:Setup}].  $\vec{E}^{'}_{far}$ and $\vec{E}^{'}_{near}$ refer to the far- and near-field contributions at the detector. The signal intensity is finally given by:

\begin{equation}
    I = \mid \vec{E}_{det} \mid^2 = \vec{E}^{'2}_{far} + 2\vec{E}^{'}_{far}\vec{E}^{'}_{near} + \vec{E}^{'2}_{near}
    \hspace{2 mm}.
    %\approx 2\vec{E}_{far}\vec{E}_{near} steht im text
    \label{eq:signal_int_no_analyzer}
\end{equation}

\noindent In order to extract the near-field contribution, typically higher-harmonic demodulation is applied to the detected signal \cite{Knoll2000}. Therefore, the tip-sample distance $h$ is modulated at the fundamental cantilever eigen-frequency $\Omega$ resulting in a modulation of the near-field signal at higher harmonics $n \Omega$ due to the strongly non-linear near-field vs. distance dependence \cite{Knoll2000}. In general $|\vec{E}^{'}_{far}| \gg |\vec{E}^{'}_{near}|$, but with increasing $n$, all far-field terms become negligible \cite{Knoll2000}. Ultimately, it is the mixed term in equation \ref{eq:signal_int_no_analyzer} that is commonly probed by this technique:

\begin{equation}
    I_{n \Omega} \approx 2\vec{E}^{'}_{far} \vec{E}^{'}_{near, n\Omega}
    = 2 (p^{'}_{far} p^{'}_{near, n\Omega} + s^{'}_{far} s^{'}_{near, n\Omega})
    \hspace{2 mm}.
    \label{eq:signal_int_no_analyzer_simplified}
\end{equation}

\noindent Here, we clearly see that, firstly, p- and s-polarized near-field contributions generally mix with their corresponding far-field component, but secondly, that p-polarization can be clearly separated from s-polarization when using polarizers and analyzers in the overall s-SNOM setup. Experimental confirmation of this finding is discussed below in section \ref{sec:illumPol}. \\

The mixing of the far and near field for any polarization is expected to be more complicated. In order to further disentangle the different vectorial near-field components, an analyzer is placed in front of the detector. The analyzer Jones matrix reads as: 

\begin{equation}
    \hat{M}_{analyzer} = 
    \begin{pmatrix}
        cos(\phi)^2   &   cos(\phi)sin(\phi) \\
        sin(\phi)cos(\phi)  &   sin(\phi)^2
    \end{pmatrix}
    \hspace{2 mm},
    \label{eq:jm_polarizer}
\end{equation}

\noindent with $\phi$ denoting the analyzer angle with respect to the p-polarization axis. The optical transmission in the detection path then becomes: 

\begin{equation}
    \begin{aligned}
        I_{analyzer} & = \mid \hat{M}_{analyzer} \cdot \vec{E}_{det} \mid^2\\
          & = cos(\phi)^2 ({p^{'}_{far}}^2 + 2 p^{'}_{far} p^{'}_{near} + {p^{'}_{near}}^2) \\
          & + sin(\phi)^2 ({s^{'}_{far}}^2 + 2 s^{'}_{far} s^{'}_{near} + {s^{'}_{near}}^2) \\
          & + \underbrace{2 cos(\phi) sin(\phi)}_{=sin(2\phi)}  (p^{'}_{far}s^{'}_{far} + \underline{p^{'}_{far}s^{'}_{near}} + \underline{p^{'}_{near}s^{'}_{far}} + p^{'}_{near}s^{'}_{near})
\hspace{2 mm}.
          \\
    \end{aligned}
    \label{eq:signal_mixing_with_analyzer}
\end{equation}

\noindent Here, the total signal intensity has 3 terms, with the first two terms describing strict p- and s-polarized states. They represent the linear polarized contributions for the trivial cases of $s_{far}=s_{near}=0$ and $p_{far}=p_{near}=0$, respectively. However, the third term accounts for the mixing of s- and p-polarized components, which intriguingly includes cross-linked terms of far and near field [see underlined terms in eq. \ref{eq:signal_mixing_with_analyzer}]. Moreover, this mixed term is no longer linearly polarized but shows a clover-like intensity distribution with orthogonal maxima at $\pm$45$^{\circ}$ due to its $2 \phi$ periodicity [see again Fig. \ref{fig:Setup_and_Permittivity} (f)]. Hence, for any incident polarization that is not strictly p- or s-polarized, a mixed polarization state can not be avoided. Extraction of these terms, again, is achieved by higher-harmonic demodulation at $n \Omega$, then reducing eq. \ref{eq:signal_mixing_with_analyzer} to:

\begin{equation}
    \begin{aligned}
        I_{analyzer} & \simeq cos(\phi)^2 \cdot 2 p^{'}_{far} p^{'}_{near, n\Omega}\\
        & + sin(\phi)^2 \cdot 2 s^{'}_{far} s^{'}_{near, n\Omega}\\
        & + sin(2\phi) \cdot (p^{'}_{far}s^{'}_{near, n\Omega} + p^{'}_{near, n\Omega}s^{'}_{far})
        \hspace{2 mm}.
    \end{aligned}
    \label{eq:signal_mixing_with_analyzer_simplified}
\end{equation}

\noindent The three components of eq. \ref{eq:signal_mixing_with_analyzer_simplified} are displayed in Figure \ref{fig:Setup_and_Permittivity} (f). Note that eqs. \ref{eq:signal_int_no_analyzer_simplified} and \ref{eq:signal_mixing_with_analyzer_simplified} form the basis for the fitting algorithm that we will apply below for data analysis, then finally allowing us to extract the sample's local permittivities. \\

%%%%%%%%%%%%%%%%%%%%%%%%%%%%%%%%%%%%%%%%%%
\section{Experimental Setup}

\begin{figure}[t]
    \centering
    \includegraphics[width = 0.7\textwidth]{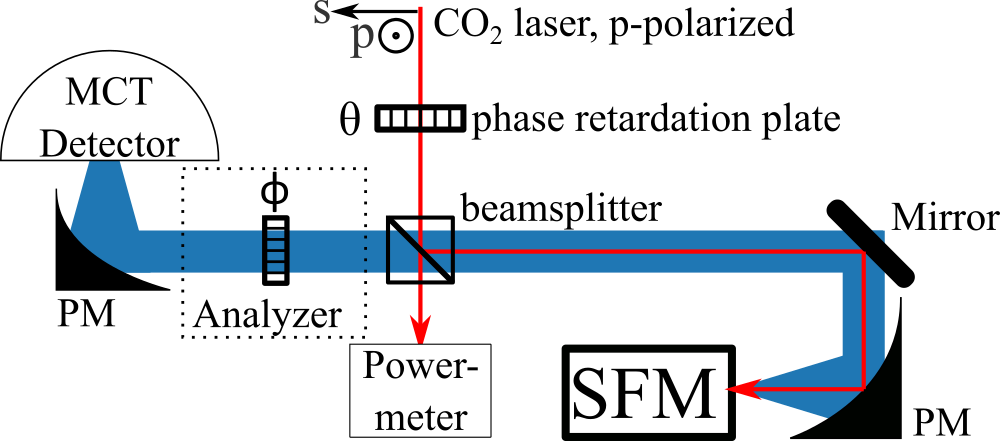}
    \caption{s-SNOM setup for the polarization-sensitive measurement scheme. Light of a CO$_2$-laser is reflected by a mylar  beamsplitter and focused via a parabolic mirror (PM) onto the SFM. The backscattered light is collected by the same PM, transmitted through the beamsplitter, and detected by a MCT detector. Illumination polarization ($\theta$) control is implemented by adding a 180$^\circ$ phase retardation plate before the beamsplitter, whereas a grid polarizer is included in the detection path to analyze the scattered signal's polarization ($\phi$).}
    \label{fig:Setup}
\end{figure}

For polarization analysis of the near-field signatures, a backscattering s-SNOM setup was adapted to include linear polarization control of both illumination and detection light paths, using a fully $360^{\circ}$ rotatable phase retardation and a grid-polarized analyzer, respectively. The setup is shown in Fig. \ref{fig:Setup}.

% components
The self-homodyne s-SNOM setup utilizes a p-polarized temperature-tunable CO$_2$ laser. Polarization control of the illuminating light is implemented by rotating the p-polarized incident state using a ZnSe 180$^\circ$ phase retardation plate, that is placed in front of the mylar beamsplitter. Note that reflection and transmission at the mylar beamsplitter itself is strongly polarization dependent, which is taken into account via the Jones matrix $\hat{M}_{mylar}$ as discussed above and in \cite{Fymat1971}. The reflected part of the laser light gets directed towards a parabolic mirror by a single gold mirror and is then focused onto the SNOM tip. Any light scattered off the tip-sample junction is then collected in backscattering geometry by the same parabola and directed towards the mylar beamsplitter via the gold mirror. The detection signal passes the mylar beam splitter and eventually also the analyzer, before then being focused onto the MCT detector via a second parabolic mirror.

All polarization measurements were performed with the s-SNOM tip kept at a fixed sample position (x,y), hence recording point-spectra. The higher-harmonic demodulation has been realized via a lock-in amplifier (integration time 50~ms), at multiples $n=2$ of the tip oscillation frequency $\Omega\approx160~kHz$. In order to investigate the role of polarization in s-SNOM, three configurations/experiments were carried out with our setup, as will be detailed in the sections to follow:

\begin{itemize}
    \item Section \ref{sec:illumPol} and Fig. \ref{fig:pol dependent intensity}:\\
    $\theta$ controlled via phase retardation plate, no analyzer in detection path;
    \item Section \ref{sec:detPol} and Fig. \ref{fig:single_lobe_plot}$:\\
    \theta=45^\circ$ set via phase retardation plate, $\phi$ probed via analyzer in detection path;
    \item Section \ref{sec:detPol} and Fig. \ref{fig:polarization_fingerprint}:\\$\theta$ controlled via phase retardation plate, $\phi$ probed via analyzer in detection path.
\end{itemize}

%%%%%%%%%%%%%%%%%%%%%%%%%%%%%%%%%%%%%%%%%%
\vspace{5 mm}
\section{Results}

\subsection{Impact of Illumination Polarization}
\label{sec:illumPol}

\begin{figure}[b]
    \centering
    \includegraphics[width=\textwidth]{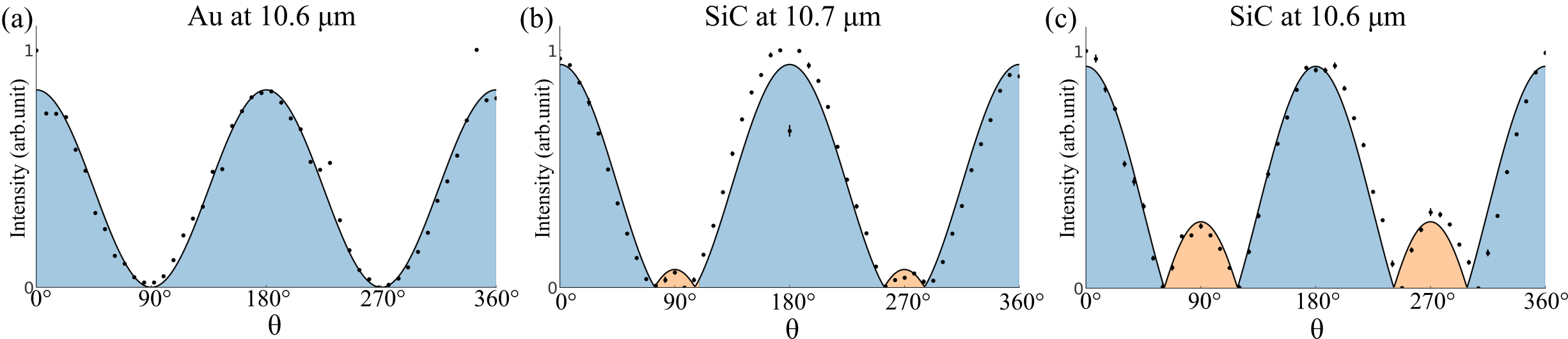}
    \caption{Second harmonic demodulated signal for rotating illumination polarization $\theta$ for different materials and illumination wavelengths. Here, $\theta_p = 0^{\circ}, 180^{\circ}, 360^{\circ}$ and $\theta_s = 90^{\circ}, 270^{\circ}$. Data are fitted using eq. \ref{eq:signal_int_no_analyzer_simplified} (black line), and p- and s-polarization-dominated contributions are shaded in \textcolor{blue}{blue} and \textcolor{orange}{orange}, respectively. (a) The gold sample illuminated with $\lambda=10.6~\mu m$ shows no response for s-polarized excitation. (b) 3C-SiC illuminated at $\lambda = 10.7~\mu m$ shows a weak response to s-polarized illumination, only, whereas, (c) an increased s-polarized response is observed when the illumination wavelength hits the sample's in-plane phonon resonance at $\lambda  = 10.6~\mu m$. All three datasets are fitted for their local sample permittivity [see tab. \ref{tab:illum_pol_fit_result}].}
    \label{fig:pol dependent intensity}
\end{figure}

As discussed above [Fig. \ref{fig:Setup_and_Permittivity} (a)], the in-plane near-field response is strongly connected to resonant excitation at certain values of the sample permittivity. The correlation between in- and out-of-plane near-field signals can therefore be directly connected to the sample's local permittivity. In the following, we thus compare the near-field response of a gold sample at 10.6~µm ($Re(\varepsilon) = -5146.8$ \cite{Weaver2015}), as a metallic non-resonant reference sample, to 3C-SiC excited resonantly at 10.6~µm ($Re(\varepsilon) = -1.40$) and 10.7~µm ($Re(\varepsilon) = -1.94$).

In order to firstly investigate the role of in-plane excitation experimentally, the illumination polarization is controlled via the 180$^\circ$ phase retardation plate [see Fig. \ref{fig:Setup}], whereas, detection is realized without using any analyzer. This corresponds to an intensity distribution as mathematically described by eq. \ref{eq:signal_int_no_analyzer_simplified}. Figure \ref{fig:pol dependent intensity} shows this intensity distribution for the second-harmonic demodulation, plotted vs. $\theta$. Each data point represents the integrated signal of  all polarization components of the scattered signal. The data sets in Fig. \ref{fig:pol dependent intensity} were then fitted using eq. \ref{eq:signal_int_no_analyzer_simplified} and normalized individually for better comparison.

For the non-resonant excitation of gold [Fig. \ref{fig:pol dependent intensity} (a)], a strong signal excited by p-polarized illumination ( $\theta_p$ = 0$^\circ$, 180$^\circ$, and 360$^\circ$) is observed together with an expected drop in intensity when rotating towards s-polarization (for $\theta_s$ = 90$^\circ$ and 270$^\circ$) [see Fig. \ref{fig:pol dependent intensity} (a)]. At $\theta_s$ the signal reduces to $I_s=0$, matching well to the dipole-model-based simulations in Fig. \ref{fig:Setup_and_Permittivity} (d). This behavior can be understood as a consequence of the boundary conditions of the dipole field at the metallic sample surface, where only electric field components perpendicular to the sample surface are allowed, hence suppressing any in-plane contributions.

This condition changes for the resonantly excited 3C-SiC sample, for which in-plane field components at the boundary are supported. Consequently, when exciting 3C-SiC at 10.7~µm [Fig. \ref{fig:pol dependent intensity} (b)], for s-polarized excitation, a non-zero near-field signal is observed, matching the theoretical prediction in Fig. \ref{fig:Setup_and_Permittivity} (d). Furthermore, when tuning the wavelength closer to the spectral resonances at 10.6 µm, the signal for s-polarized excitation increases [see Fig. \ref{fig:pol dependent intensity} (c)] to about 25\% of the intensity at p-polarized excitation. This clearly underlines that in-plane near-field excitation indeed is possible leading to a quite strong and measurable s-SNOM signal. Furthermore, the intensity of the in-plane signal correlates with the resonance as predicted by the tip-dipole model.

In order to fit the data, eq. \ref{eq:signal_int_no_analyzer_simplified} was used, leading to a qualitatively correct result for all three plots in Figure \ref{fig:pol dependent intensity}. As result of fitting, the sample permittivities could be extracted [see tab. \ref{tab:illum_pol_fit_result}], illustrating two findings: Firstly, the highly non-resonant nature of gold at 10.6~µm can be reproduced by the fit. Secondly, the comparably small difference in sample permittivity for resonant excitation could be resolved successfully for 3C-SiC at 10.6~µm and 10.7~µm, respectively. The deviation between the values determined by us and the literature values, is likely caused by inaccurate fitting for a single polarization line. In particular,
%. In order to retrieve more exact data, the signals polarization state needs to investigated in addition.
as detailed in section \ref{sec:Theory}, the near-field interaction depends on e.g. tip radius, modulation amplitude, and tip-sample distance in addition to sample permittivity. These important system parameters are not fitted reliably by the single-line measurement and need to be taken into account. To increase the accuracy further and thereby enable a more detailed polarization examination, in the following an analyzer will be added to the detection path.

\begin{table}[t] 
\caption{$Re(\varepsilon)$ for Au and 3C-SiC as extracted through fitting the experimental results shown in Figure~\ref{fig:pol dependent intensity} by applying eq. \ref{eq:signal_int_no_analyzer_simplified}.}
\newcolumntype{C}{>{\centering\arraybackslash}X}
\begin{tabularx}{\textwidth}{CCC}
\toprule
\textbf{sample}	& \textbf{Re($\varepsilon$); our result}	& \textbf{Re($\varepsilon$); literature}\\
\midrule
         Au @ 10.6 µm  & -6521.02 & -5146.8 \cite{Weaver2015} \\
         3C-SiC @ 10.7 µm & -18.23 $\pm$  1.55   & -1.94 \cite{Mutschke1999} \\
         3C-SiC @ 10.6 µm & -4.27  $\pm$  0.53  & -1.40 \cite{Mutschke1999} \\
\bottomrule
\label{tab:illum_pol_fit_result}
\end{tabularx}
\end{table}

\subsection{Analyzing the Near-Field Polarization}
\label{sec:detPol}

\begin{figure*}[t]
    \centering
    \includegraphics[width=\textwidth]{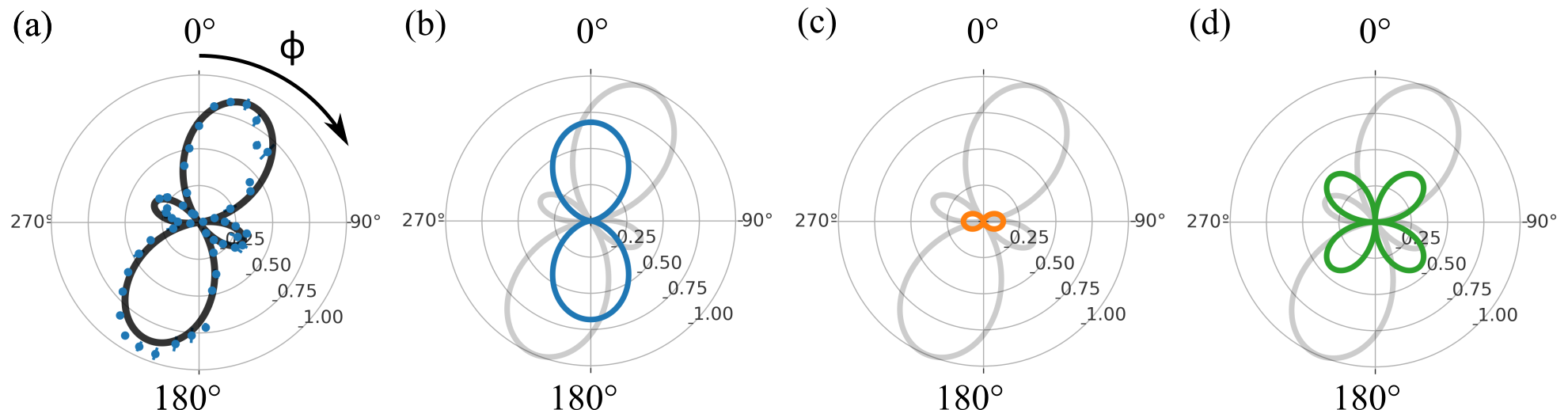}
    \caption{Polarization distribution of the near-field signal for a 3C-SiC sample measured as a function of analyzer rotation $\phi$ ($\lambda$ = 10.6~µm, $\theta$ = 45$^\circ$). (a) Overall distribution and fit; (b) p-polarized, (c) s-polarized, and (d) mixed contributions, calculated and extracted from the fit showing a strong coupling even for small s-polarized contributions due to component mixing [see eq.~\ref{eq:signal_mixing_with_analyzer_simplified}]. For comparison, the fitted overall distribution from (a) is reproduced in gray in (b-d).}
    \label{fig:single_lobe_plot}
\end{figure*}

In addition to measuring the integrated signal at a defined illumination polarization, we furthermore investigate the nature of complex far- and near-field mixing by analyzing the polarization distribution of the scattered light. Therefore, the different contributions in the backscattered signal are separated by an analyzer at the angle $\phi$ with respect to p-polarization.

As an exemplary measurement we excite 3C-SiC resonantly at 10.6~µm with $\theta = 45^\circ$-polarized incident light [Fig. \ref{fig:single_lobe_plot}]. The overall signal [see Fig. \ref{fig:single_lobe_plot} (a)] shows a non-trivial polarization state with four lobes, the maxima of which are 90$^\circ$ apart. The stronger lobe is oriented along the 25$^\circ$ axis, leaving an offset of $\sim 20^\circ$ with respect to the illumination polarization. In order to extract the underlying polarization contributions, the lobe is fitted using eq. \ref{eq:signal_mixing_with_analyzer} resulting in an excellent match. Moreover, the fit allows for a separation of the underlined terms and hence for the extraction of individual signal contributions. The corresponding separated p-polarized, s-polarized, and mixed terms are shown in Fig. \ref{fig:single_lobe_plot} (b), (c), and (d), respectively, with the overall signal in grey for comparison. Comparing p- and s-polarized contributions, the out-of-plane field enhancement of the former is apparent, yielding a five times higher signal intensity. However, this comparably small s-polarized contribution causes a significant rotation of the overall light distribution due to the mixed term. Referring to eq. \ref{eq:signal_mixing_with_analyzer}, the origin of this enhancement is found in the mixing of the p-polarized far-field term with the s-polarized near-field term and vice versa, resulting in the clover-like polarization response as displayed in Fig. \ref{fig:single_lobe_plot} (d).

\begin{figure*}[b]
    \centering
    \includegraphics[width=0.8\textwidth]{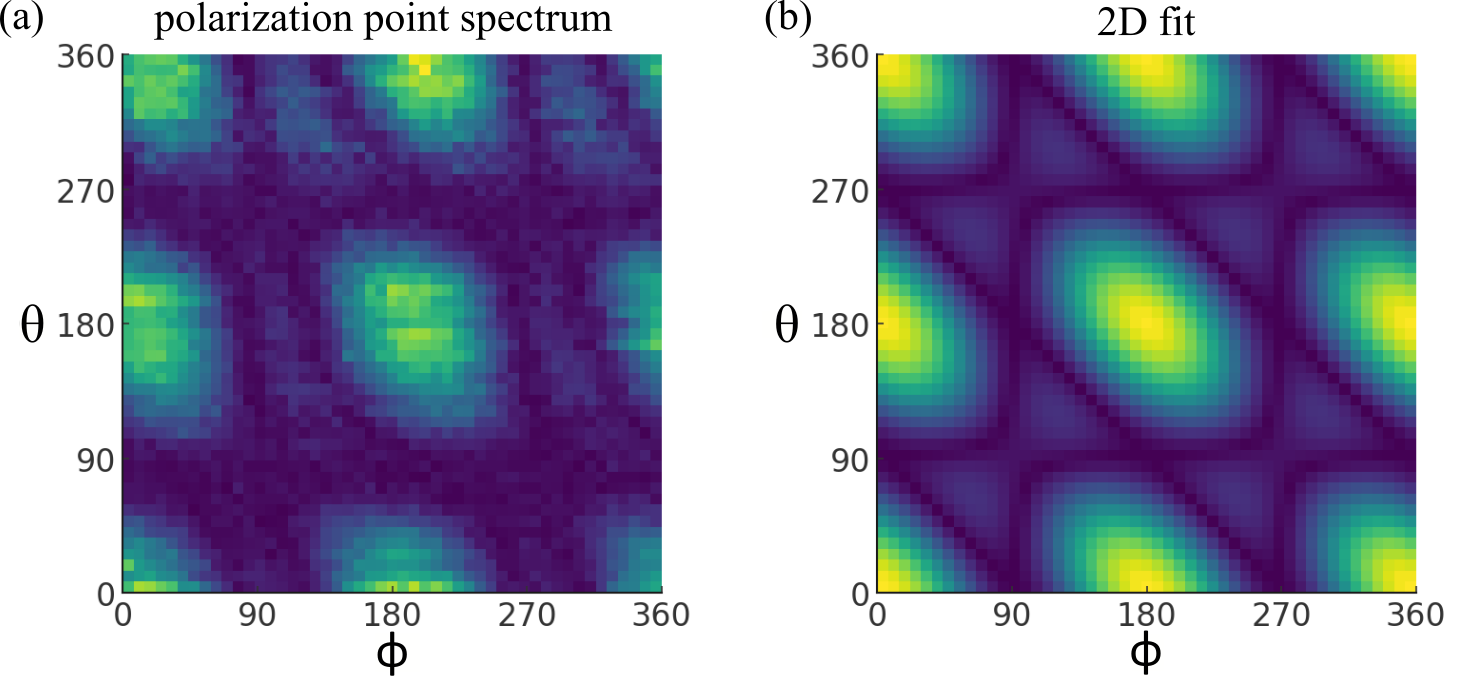}
    \caption{Normalized polarization point spectrum of 3C-SiC at 10.6~µm. (a) The polarization point spectrum was measured by rotating the analyzer ($\phi$) by 360$^\circ$ for each illumination polarization angle ($\theta$). (b) The 2D fit of the polarization point spectra matches well with the experimental results and is used to extract the local sample permittivity (see Table \ref{tab:pol_footprint_fit_result}).}
    \label{fig:polarization_fingerprint}
\end{figure*}

~\\
In order to explore the origin of the complex polarization mixing, we extend our measurements to a full set of incident and scattered polarizations as displayed in Fig. \ref{fig:polarization_fingerprint}. Here, the intensity distribution in each horizontal line represents one full rotation of the analyzer angle ($\phi$) while the vertical axis corresponds to varying incident polarization angles $\theta$. We refer to the resulting 2D plot as "polarization point spectrum".

Due to the symmetry of the underlying polarization, the polarization point spectrum [see Fig. \ref{fig:polarization_fingerprint}] has a twofold symmetry (and hence self-reproduces every 180$^\circ$). Furthermore, the strongest signal is observed for p-polarized illumination, for $\theta = 0^\circ, 180^\circ, 360^\circ$ and p-polarized detection at $\phi = 0^\circ, 180^\circ, 360^\circ$, as expected when comparing it to the integrated intensity distribution in Fig. \ref{fig:pol dependent intensity} (c). Moreover, sidemaxima are observed for s-polarized analyzer angles at $\phi\approx90^\circ$.

In order to fit the data, eq. \ref{eq:signal_mixing_with_analyzer} is used, in analogy to the fit for Fig. \ref{fig:single_lobe_plot} (a) as described above. The function was modified, however, to fit the entire polarization point spectrum recursively. Qualitatively, the resulting fit resembles the experimental data very accurately, including both the global maxima at $\theta$, $\phi=0^\circ, 180^\circ$ and $360^\circ$, as well as the corresponding sidemaxima. The latter resembles the creation of side lobes as connected to the mixing of far- and near-field components with different polarization states (see eq.~\ref{eq:signal_mixing_with_analyzer_simplified}).  
%Most prominently, the tip-polarizability and oscillation amplitude have been included in addition to the sample permittivity for the recursive fitting algorithm. All three values are assumed to be independent of the incident polarization state, based on the theory formulated in section \ref{sec:Theory}. By defining the bounds of the parameters to match real-life values (e.g. limiting the oscillation amplitude to an interval of 40 - 200 nm), a good fit for the permittivity was achieved as summarized in Table \ref{tab:pol_footprint_fit_result}.
% what parameters
Quantitatively, the recursive fit enables the determination of the sample's permittivities at two different wavelengths as summarized in Table \ref{tab:pol_footprint_fit_result} which are close, but not perfectly matching to the values found in literature. Deviations might be related to the here-assumed spherical tip in comparison to more realistic tip shapes \cite{McArdle2020}, which will be in the focus of future work. However, compared to the investigation of incident polarization only as summarized in table \ref{tab:illum_pol_fit_result}, the advanced polarization-point-spectrum fit improved the obtained Re($\varepsilon$) values significantly, demonstrating the relevance of a complete mapping of near-field signal polarization.

Overall, the successful extraction of local permittivity data from  polarization point spectra demonstrates the potential of polarization-controlled s-SNOM for the local material characterization.

% add other data and error
\begin{table}[t] 
\caption{Re($\varepsilon$) for 3C-SiC evaluated at different wavelengths, resulting from fitting the polarization point spectra using eq. \ref{eq:signal_mixing_with_analyzer}.}
\label{tab:pol_footprint_fit_result}
\newcolumntype{C}{>{\centering\arraybackslash}X}
\begin{tabularx}{\textwidth}{CCCCC}
\toprule
\textbf{wavelength}	& \textbf{Re($\varepsilon$); our results}   & \textbf{Re($\varepsilon$); literature} \\
\midrule
10.6 µm		& -2.57 $\pm$ 0.54   & -1.40 \cite{Mutschke1999}\\
10.7 µm		& -6.37 $\pm$ 0.73   & -1.94 \cite{Mutschke1999}\\
\bottomrule
\end{tabularx}
\end{table}

%%%%%%%%%%%%%%%%%%%%%%%%%%%%%%%%%%%%%%%%%%
\vspace{5 mm}

\section{Conclusions}

In summary, we demonstrated that resonant sample excitation in s-SNOM results in significant in-plane contributions that can be addressed individually by adjusting the illumination polarization angle accordingly. The in-plane response is strongly connected to the sample permittivity as demonstrated for resonantly and non-resonantly excited samples. Moreover, the mixing between s- and p-polarized far- and near-field components was demonstrated to lead to a rotation in the scattered polarization, even for small s-polarized contributions. Finally, full polarization analysis enables the extraction of the local permittivity data via a recursive fit algorithm that matches well to the values reported in literature.

More generally, by demonstrating the existence and reading of in-plane near-field components, this fundametnal work forms an important pillar for future polarization-sensitive s-SNOM application scenarios, such as nanoellipsometry e.g. by introducing also circularly-polarized field excitation. Particularly, we envision the application towards in-plane sensitive and anisotropic samples such as 2D van-der-Waals and topological materials.

%%%%%%%%%%%%%%%%%%%%%%%%%%%%%%%%%%%%%%%%%%
\begin{adjustwidth}{-\extralength}{0cm}

%%%%%%%%%%%%%%%%%%%%%%%%%%%%%%%%%%%%%%%%%%
%For research articles with several authors, a short paragraph specifying their individual contributions must be provided. The following statements should be used.
\vspace{10 mm}
\authorcontributions{Conceptualization, S.C. Kehr and F.G. Kaps; methodology, S.C. Kehr and F.G.Kaps; validation, F.G. Kaps and S.C. Kehr; formal analysis, F.G. Kaps.; investigation, F.G. Kaps; resources, S.C. Kehr and L.M. Eng; data curation, F.G. Kaps; writing---original draft preparation, F. G. Kaps; writing---substantial review and editing, S.C. Kehr and L.M. Eng; visualization, F.G. Kaps; supervision, S.C. Kehr and L.M. Eng; project administration, S.C. Kehr and L.M. Eng; funding acquisition, S.C. Kehr and L.M. Eng. All authors have read and agreed to the published version of the manuscript.}

\funding{This research was funded by Bundesministerium für Bildung und Forschung (BMBF) grant numbers 05K19ODA, 05K19ODB, 05K22ODA and the Deutsche Forschungsgemeinschaft (DFG) through project \mbox{CRC1415 (ID: 417590517)} as well as the Würzburg-Dresden Cluster of Excellence” ct.qmat” (EXC 2147).}

\acknowledgments{The authors thank R. Buschbeck for support in optimizing and structuring the simulations.}

\conflictsofinterest{The authors declare no conflict of interest.} 

\reftitle{References}

\bibliography{main.bib}

\end{adjustwidth}
\end{document}